\documentclass{moriond}

\usepackage{amsmath}
\usepackage{amssymb}
\usepackage{bm}
\usepackage{amssymb}

\def\be{\begin{equation}}
\def\ee{\end{equation}}
\def\bea{\begin{eqnarray}}
\def\eea{\end{eqnarray}}

\newcommand{\Photo}{}

\begin{document}
\vspace*{4cm}
\title{UNIVERSALITY OF FREE FALL VERSUS EPHEMERIS}

\author{O. Minazzoli \dag \ddag\\
L. Bernus$^{\alpha,\beta}$, A. Fienga$^\beta$, A. Hees$^\gamma$, J. Laskar$^\alpha$, V. Viswanathan $^\beta$ }

\address{\dag Centre Scientifique de Monaco, 8 quai Antoine I, Monaco\\ \ddag Artemis, Universit\'e C\^ote d'Azur, Observatoire C\^ote d'Azur, CNRS, CS 34229, F-06304 Nice, France\\
$^\alpha$ ASD/IMCCE, CNRS-UMR 8028, Observatoire de Paris, PSL, UPMC, 77 avenue Denfert-Rochereau, 75014 Paris, France\\ 
$^\beta$ G\'eoAzur, CNRS-UMR 7329, Observatoire de la Côte d’Azur, Université Nice Sophia Antipolis, 250 Av. A. Einstein,
06560 Valbonne, France
\\  $^\gamma$ Department of Physics and Astronomy, University of California, Los Angeles, CA 90095, USA}

\maketitle\abstracts{
When a light scalar field with gravitational strength interacts with matter, the weak equivalence principle is in general violated, leading for instance to a violation of the universality of free fall. This has been known and tested for a while. However, recent developments [Minazzoli \& Hees, PRD 2016] showed that a novel manifestation of the universality of free fall can appear in some models. Here we discuss this new scenario and expose how we intend to constrain it with INPOP ephemeris.}

\section{Introduction}

Massless or light scalar-fields with gravitational strength that directly couple to matter are expected in the context of string theory \cite{damour:1994np,damour:2010pr}. A consequence of this type of fields would be that the Equivalence Principle (EP) is violated  \cite{damour:1994np,damour:2010pr}. More precisely, the Einstein Equivalence Principle (EEP) would be violated, with several manifestations, such as violations of the Local Position Invariance (LPI) as well as violations of the Universality of Free Fall (UFF) --- also known as Weak Equivalence Principle (WEP).

The most precise tests of the UFF have been made by comparing the free fall accelerations of different test bodies \cite{schlamminger:2008pl}. It is usually thought that the relative acceleration (at the Newtonian level) between  two bodies that are equidistant from the source of gravity reads as follows \cite{williams:2012cq}
\begin{equation}
\frac{\Delta a}{a} \equiv 2 \frac{a_1-a_2}{a_1+a_2} = \left[\frac{m^G}{m^I} \right]_1-\left[\frac{m^G}{m^I} \right]_2= \Delta\left[\frac{m^G}{m^I} \right],
\end{equation}
where $m^G$ and $m^I$ are the gravitational and inertial masses of each body respectively. However, recent phenomenological developments suggest that it may actually be more complicated than that in some situations \cite{hees:2015ax,minazzoli:2016pr}, as we shall see bellow. In any case, the planetary and lunar ephemeris INPOP \cite{fienga:2008aa,fienga:2011cm} is an ideal tool in order to implement EP tests.

\section{Brief description of INPOP}

INPOP (Int\'{e}grateur Num\'{e}rique Plan\'{e}taire de l'Observatoire de Paris) is a planetary ephemeris that is built by integrating numerically the equations of motion of the solar system, and by adjusting to lunar laser ranging and space missions' observations \cite{fienga:2008aa,fienga:2011cm,fienga15}. In addition to the classic planetary and lunar fitted parameters, one can add parameters encoding deviations from general relativity. These parameters can be adjusted simultaneously with all the others in a global fit. With this method, good constraints were put on the PPN parameters \cite{fienga15} --- using Mercury orbiter data (MESSENGER) \cite{verma:2014aa}, but also by considering a Monte Carlo exploration of the solutions' space~\cite{fienga15}. The same methods can be used for adjusting the new parameters described in this work.

\section{Acceleration at the Newtonian level}

Considering a general scalar-tensor theory with non-minimal scalar-matter coupling (that cannot be gauged away by a metric redefinition such as a conformal or a disformal transformation), it has recently been shown that the acceleration of a body, say $T$, reads \cite{hees:2015ax,minazzoli:2016pr}
\begin{equation}
\bm {a}_T=-\sum_{A\neq T} \frac{G m^G_A}{r_{AT}^3}\bm r_{AT}\left(1+\delta_T+\delta_{AT}\right) \label{eq:EIH_s},
\end{equation}
where $\bm r_{AT}= \bm x_T - \bm x_A$. The coefficients $\delta_T$ and $\delta_{AT}$ parametrize the violation of the UFF.  $G$ is the ``measured'' constant of Newton and $m^G_A$ is the ``gravitational'' mass of the body $A$. It is important to have in mind that $G$ and $m^G_A$ are not the constant of Newton and the mass that appear in the fundamental action \cite{hees:2015ax,minazzoli:2016pr}. One notably has $m^G_A=(1+\delta_A) m^I_A$, where $m^I_A$ is the inertial mass of the body $A$ \cite{hees:2015ax,minazzoli:2016pr}. As a consequence, from equation (\ref{eq:EIH_s}), one can check that the gravitational force in this context still satisfies Newton's third law of motion:
\begin{equation}
m^I_A\bm {a}_A =\frac{G m^I_A m^I_B}{r_{AB}^3}\bm r_{AB}\left(1+\delta_A+\delta_B+\delta_{AB}\right)=- m^I_B\bm {a}_B.
\end{equation}
In general, $\delta_T$ can be decomposed into two contributions: one from a violation of the WEP and one from a violation of the Strong Equivalence Principle (SEP):
\begin{equation}
\delta_T=\delta_T^{WEP}+\delta_T^{SEP},\qquad \textrm{where}\qquad \delta_T^{SEP} = \eta \frac{|\Omega_T|}{m_T~c^2},
\end{equation}
where $\Omega$ and $m c^2$ are the gravitational binding and rest mass energies respectively, while $\eta$ is the so-called Nordtvedt parameter. On the other side, $\delta_T^{WEP}$ depends on both the scalar-matter coupling parameters and on the dilatonic charges \cite{damour:2010pr,hees:2015ax,minazzoli:2016pr}. In most cases, if $\delta_T^{WEP} \neq 0$, then $\delta_T^{WEP} \gg  \delta_T^{SEP}$, such that one can usually test either the WEP (discarding SEP violations), or the SEP (discarding WEP violations).

As the parameter $\delta_T^{WEP}$, $\delta_{AT}$ depends on both the scalar-matter coupling parameters and on the dilatonic charges \cite{hees:2015ax,minazzoli:2016pr}. In most situations, $\delta_T^{WEP} \gg \delta_{AT}$. However, it is not necessarily true when the scalar-matter coupling is the same in each sector of particle physics. In that situation, one can have $\delta_T^{WEP} \lesssim \delta_{AT} $ \cite{hees:2015ax,minazzoli:2016pr}. It is noteworthy  that such kind of universality has already been suggested in the context of string theory \cite{damour:1994np}.

The important thing to notice with $\delta_{AT}$, is that it depends not only on the composition of the falling body, but also on the composition of the body that is source of the gravitational field in which the body $T$ is falling. As a consequence, the relative acceleration of two test particles cannot only be related to the ratios between their gravitational to inertial masses.

\section{The Earth-Moon system}

At the Newtonian level, the relative acceleration between the Earth and the Moon reads
\begin{equation}
\bm a_M - \bm a_E= -  \frac{ G \mu}{r_{EM}^3}\bm r_{EM}+   G m^G_S\left[ \frac{\bm r_{SE}}{r_{SE}^3}-\frac{\bm r_{SM}}{r_{SM}^3} \right]+   G  m^G_S\left[ \frac{\bm r_{SE}}{r_{SE}^3} (\delta_E+\delta_{SE})- \frac{\bm r_{SM}}{r_{SM}^3} (\delta_M+\delta_{SM}) \right], \label{eq:EMacc}
\end{equation}
With $\mu\equiv  m^G_M +  m^G_E+ (\delta_E+\delta_{EM})m^G_M+(\delta_M+\delta_{EM}) m^G_E$. 
With ephemeris, the first term of equation (\ref{eq:EMacc}) does not lead to a sensitive test of the UFF, because it can be absorbed in the fit of the parameter $m^G_M+m^G_E$.\cite{williams:2012cq} The last term, on the other side, does. At leading order, one can approximate both distances appearing in this last term as being approximately equal. One therefore has
\begin{eqnarray}
\Delta \bm a^{\bar UFF}\equiv (\bm a_M - \bm a_E)^{\bar UFF} &\approx&  G m^G_S\left[ \frac{\bm r_{SE}}{r_{SE}^3} (\delta_E+\delta_{SE})- \frac{\bm r_{SM}}{r_{SM}^3} (\delta_M+\delta_{SM}) \right], \nonumber \\
&\approx&\bm a_{E}  \left[(\delta_E+\delta_{SE})- (\delta_M+\delta_{SM})\right],
\end{eqnarray}
where  $\Delta \bm a^{\bar UFF}$ is the part of the relative acceleration between the Earth and the Moon that violates the UFF. When $\delta_{SM}=\delta_{SE}$, one recovers the usual expectation, that is \cite{williams:2012cq}
\begin{equation}
\Delta \bm a^{\bar UFF} \approx \bm a_{E} \left[\left(\frac{m^G}{m^I} \right)_E-  \left(\frac{m^G}{m^I} \right)_M \right].
\end{equation}

The results from the comparison of the numerical integration of Eq. (\ref{eq:EIH_s}) to the measurement of the Earth-Moon distance via Lunar Laser Ranging will be published in a dedicated communication.

As one can see, there are more parameters than equations of motion. Therefore, the Earth-Moon system alone constrain a specific combination of these parameters only. In consequence, it may be useful to take advantage of the many bodies that are in the solar system.

\section{Planetary orbits}

One can show that the parameters $\delta_A$ and $\delta_{AT}$ mostly depend on six fundamental (or semi-fundamental) parameters --- related to the couplings between the scalar field and each sector of particle physics \cite{damour:2010pr,hees:2015ax,minazzoli:2016pr}. As a consequence, in order to constrain those parameters --- naively --- one needs to observe at least 6 falling bodies, with sensible different compositions. Therefore, one may use solar system observations in order to constrain those parameters individually --- although with a weaker accuracy than what can be achieved with the Earth-Moon system alone.

\section{Acceleration at the post-Newtonian level}

At current level accuracy for Solar system observations (e.g. $\sim 1cm/20yrs$ for the Moon and $\sim 1m/20yrs$ for Mars), one has to deal with the full post-Newtonian (pN) equation of motion. In the present context, it reads \cite{hees:2015ax}
\begin{align}
	\bm {a}_T=&-\sum_{A\neq T} \frac{ G m^G_A}{r_{AT}^3}\bm r_{AT}\left(1+\delta_T+\delta_{AT}\right) \label{eq:EIH} \\
	&-\sum_{A\neq T} \frac{ G m^G_A}{r_{AT}^3c^2}\bm r_{AT}\Bigg\{\gamma v_T^2+(\gamma+1)v_A^2-2(1+\gamma)\bm v_A.\bm v_T  -\frac{3}{2}\left(\frac{\bm r_{AT}.\bm v_A}{r_{AT}}\right)-\frac{1}{2}\bm r_{AT}.\bm a_A \nonumber \\
	&\hspace{3cm}-2(\gamma+\beta+\mathrm{d}\beta^T)\sum_{B\neq T}\frac{ G m^G_B}{r_{TB}}-(2\beta+2\mathrm{d}\beta^A-1)\sum_{B\neq A}\frac{ G m^G_B}{r_{AB}}\Bigg\} \nonumber \\
	&+\sum_{A\neq T}\frac{ G m^G_A}{c^2r_{AT}^3}\left[2(1+\gamma)\bm r_{AT}.\bm v_T-(1+2\gamma)\bm r_{AT}.\bm v_A\right](\bm v_T-\bm v_A) + \frac{3+4 \gamma}{2}\sum_{A\neq T} \frac{ G  m^G_A}{c^2r_{AT}}\bm a_A\nonumber \, ,
\end{align}
where $\gamma$ and $\beta$ are the usual pN parameters. $\mathrm{d}\beta^X$ is a new parameter that depends on the composition of the body $X$. It indicates how non-linear is the scalar-matter coupling \cite{hees:2015ax}. We do not expect that it plays a significant role in the pN dynamics though \cite{hees:2015ax}. All those parameters can be expressed in terms of some of the fundamental parameters discussed in the previous section~\cite{hees:2015ax,minazzoli:2016pr}. 
\section{Conclusion}

The solar system is ideal in order to constrain possible non-minimal couplings between gravity and matter. Here we discussed one such potential couplings, where one gravitational (massless or light) scalar field couples multiplicatively with different sectors of particle physics. Thanks to current and future advances in the solar system exploration, one can expect to greatly increase the accuracy of solar system EP tests in a foreseeable future. Based on the equations of motion presented here, the INPOP software developed at both the Paris and Nice observatories will allow such tests.




%
%
%
%

\end{document}